\documentclass[twocolumn,showpacs,preprintnumbers,amsmath,amssymb]{revtex4-1}
\usepackage{graphicx}
\usepackage{textcomp}

\begin{document}
%\draft
\title{Classical simulation of the Hubbard-Holstein dynamics with optical waveguide lattices}
  \normalsize
\author{Stefano Longhi and Giuseppe Della Valle}
\address{Dipartimento di Fisica, Politecnico di Milano, Piazza L. da Vinci
32, I-20133 Milano, Italy}

%\date{.}
%
\bigskip
\begin{abstract}
A classical analog simulator of the two-site Hubbard-Holstein model,
describing the dynamics of two correlated electrons coupled with
local phonons, is proposed based on light transport in engineered
optical waveguide arrays. Our photonic analog simulator enables to
map the temporal dynamics of the quantum system in Fock space  into
spatial propagation of classical light waves in the
evanescently-coupled waveguides of the array.  In particular, in the
strong correlation regime the periodic temporal dynamics, related to
the excitation of Holstein polarons with equal energy spacing,
can be visualized as a periodic self-imaging phenomenon of the light
beam along the waveguide array and explained in terms of generalized Bloch oscillations of a single particle 
in a semi-infinite inhomogeneous tight-binding lattice. 
 \noindent
\end{abstract}


% 71.10.Fd Lattice fermion models (Hubbard model, etc.)
% 71.38.-k Polarons and electron-phonon interactions 
% 71.38.Ht Self-trapped or small polarons
% 42.82.Et Waveguides, couplers, and arrays

\maketitle

\section{Introduction}
In recent years, there has been a renewed and increasing interest
in the development of quantum analog simulators, i.e.
controllable quantum systems that
can be used to imitate other quantum systems (see, for instance, \cite{Nori} and references therein).
Quantum analog simulators would be very useful for a
wide variety of problems in physics, chemistry,
and biology. In particular, they have proven to be able to tackle complex quantum many-body
problems in condensed-matter physics \cite{Lewenstein,Bloch,i0}, such as correlated
electrons or quantum magnetism. Quantum analog simulators are also useful to test predictions  or states of matter difficult to access experimentally in the originally proposed system, such as dynamical and extreme conditions \cite{Nori}. 
Recent realizations of
quantum analog simulators of many-body problems have been  based mainly on atoms \cite{Lewenstein,a1,a2,a3}, trapped ions \cite{i0,i1,i2,i3}, nuclear magnetic resonance \cite{mr1}, and single photons \cite{ppp1,ppp2}. As compared to quantum simulators based on atoms or ions, those employing single photons offer the feasibility to
 individually address and access the dynamics of single particles, and are thus suited to simulate the physics of few interacting particles \cite{ppp2}.
 Unlike electrons, photons do not interact with each other, especially at the few photon level, and hence for the simulation of  quantum many-body
problems of condensed-matter physics particle interaction has to be introduced in a suitable manner.
  At the few photon level,  interaction is generally mimicked using purely linear optical systems by measurement-induced nonlinearity \cite{Milburn}.
  In this way, simplified toy models of many-body condensed matter physics have been simulated, such as frustrated Heisenberg dynamics in a spin-1/2 tetramer \cite{ppp2}.
   On the other hand, propagation of classical light in waveguide-based optical lattices has provided an experimentally accessible test bed to mimic in a purely
  classical setting {\em single-particle} coherent phenomena of solid-state physics \cite{LonghiLPR,Szameit10}, such as Bloch oscillations \cite{BO}, dynamic localization \cite{DL}, and Anderson localization \cite{AL} to mention a few.    Recently, the possibility to simulate Bose-Hubbard and Fermi-Hubbard models of  few interacting bosons or fermions in coupled-waveguide {\em linear} optical structures  using {\it classical}  light beams has been proposed as well \cite{Longhi1,Longhi2}. The main idea is that the temporal evolution of a
 few interacting particles in Fock space can be conveniently mapped into {\em linear} spatial propagation of discretized light \cite{Lederer} in engineered evanescently-coupled optical waveguides.  As compared to {\em quantum simulators} using single photons, atoms or trapped ions, such a kind of {\em classical simulators} of quantum many-body problems using classical light would benefit for the much simpler experimental implementation, the rather unique possibility to
provide a direct access and visualization of the temporal dynamics of the quantum system in Fock space, and the ability to prepare the system in a rather arbitrary (for example highly entangled) state. \\
In condensed-matter, strong correlation effects and localization can occur in metallic systems due both to strong
electron-electron({\it e-e})  interactions and strong electron-phonon ({\it e-ph}) coupling. Evidences of strong {\it e-ph} interactions
have been reported in such important materials as cuprates, fullerides, and manganites (see, e.g., \cite{cuprates} and references therein).
The interplay of {\it e-e} and {\it e-ph} interactions
in these correlated systems leads to coexistence of
or competition between various phases such as superconductivity,
charge-density-wave or spin-density-wave  phases, or formation
of novel non-Fermi liquid phases, polarons, bipolarons,
etc.  One of the simplest theoretical model that accounts for the interplay between these two types of interactions is the Holstein-Hubbard (HH) model \cite{HH0,HH1,HH2}.
Even though the HH model provides an oversimplified description of
both {\it e-e} and {\it e-ph} interactions, it retains what are thought to be the relevant ingredients of
a system in which electrons experience simultaneously an instantaneous short-range repulsion
and a phonon-mediated retarded attraction. Actually, in spite of its formal simplicity, it is
not exactly solvable even in one dimension. As quantum simulators of Fermi-Hubbard models, based on either fermionic atoms in optical lattices \cite{Bloch,FHP} or electrons in artificial quantum-dot crystals    \cite{CQD}, have attracted great interest, analog simulators of HH models have not received special attention yet. The smallest configuration for studying and simulating the HH dynamics is the dimer HH model, which  describes two correlated electrons
 hopping between two adjacent sites \cite{HH2}. In In this work, a classical analog simulator is theoretically proposed for the two-site HH model, based on light transport in engineered waveguide lattices, which is capable of reproducing the temporal dynamics of the quantum model in Fock space as spatial beam evolution in the photonic lattice.
The paper is organized as follows. In Sec.II, the HH model is briefly reviewed, and the dynamical equations of Fock-state amplitudes are derived.
In Sec.III, a classical analog simulator based on light transport in engineered waveguide arrays  is described, and numerical results are presented for waveguide structures that simulate the HH model for different electron correlation regimes. In particular,
in the strong correlation regime the quasi-periodic temporal dynamics, related to the excitation of
Holstein polarons, is  visualized as a periodic self-imaging
phenomenon of the light beam that propagates along the waveguide array and explained in terms of 
generalized Bloch oscillations of a single particle in an inhomogeneous biased tight-binding lattice. Finally, in Sec. IV the main conclusions are outlined.

\section{The Hubbard-Holstein model: Fock-space representation}
\subsection{The two-site Hubbard-Holstein model}
The two-site HH model describes two correlated
electrons hopping between two adjacent sites (a diatomic molecule),
each of which exhibits an optical mode with frequency
$\Omega$. This model has been often used in condensed-matter physics as a simplified toy model 
to describe some basic features of polaron and bipolaron dynamics (see, for instance, \cite{HH2}). 
The two-site HH Hamiltonian can be separated into two
terms \cite{HH2}. One describes a shifted oscillator that does not
couple to the electronic degrees of freedom, and can be thus disregarded. The other describes
the effective {\it e-ph} system where phonons couple directly
with the electronic degrees of freedom. The corresponding Hamiltonian, with   $\hbar=1$, reads explicitly \cite{HH2}
\begin{equation}
\hat{H}=\hat{H}_{e}+\hat{H}_{ph}+\hat{V}_{e-ph}
\end{equation}
where
\begin{eqnarray}
\hat{H}_{e} & = & -t \sum_{\sigma=\uparrow, \downarrow} \left( \hat{c}^{\dag}_{1,\sigma}\hat{c}_{2,\sigma}+\hat{c}^{\dag}_{2,\sigma}\hat{c}_{1,\sigma}  \right)+ \nonumber \\
& + & U \left( \hat{n}_{1,\uparrow} \hat{n}_{1,\downarrow} +   \hat{n}_{2,\uparrow} \hat{n}_{2,\downarrow}  \right)   \\
\hat{H}_{ph} & = & \Omega \hat{b}^{\dag}\hat{b} \\
\hat{V}_{e-ph} & = & \frac{g}{\sqrt 2} \left( \hat{n}_1-\hat{n}_2 \right) \left( \hat{b} + \hat{b}^{\dag} \right).
\end{eqnarray}
In the previous equations, $\hat{c}^{\dag}_{i, \sigma}$ ($\hat{c}_{i, \sigma}$) are the fermionic creation (annihilation)
operators for the electron at site $i$ with spin $\sigma$ ($i=1,2$, $\sigma= \uparrow, \downarrow$),
$\hat{n}_{i , \sigma}=\hat{c}^{\dag}_{i, \sigma}\hat{c}_{i, \sigma}$ and $\hat{n}_i=\hat{n}_{i, \uparrow}+\hat{n}_{i, \downarrow}$
are the electron occupation numbers,  $g$ is the on-site {\it e-ph}
coupling strength, $t$ is the hopping amplitude between adjacent sites, $U$ is the on-site Coulomb interaction energy,
and $\hat{b}^{\dag}$ ($\hat{b}$)
 is creation (annihilation) bosonic operator for the oscillator.
For the two-site two-electron system there are six electronic
states. Three of these states are degenerate, with zero energy $\hat{H}_{e} |\psi_{Ti} \rangle=0$ ($i=1,2,3$), and belong to the triplet states as follows:
\begin{eqnarray}
| \psi_{T1} \rangle & = & \hat{c}^{\dag}_{1, \uparrow} \hat{c}^{\dag}_{2,\uparrow} |0 \rangle_{e} \\
| \psi_{T1} \rangle & = & \frac{1}{\sqrt{2}} \left(  \hat{c}^{\dag}_{1, \uparrow} \hat{c}^{\dag}_{2,\downarrow}+\hat{c}^{\dag}_{1, \downarrow} \hat{c}^{\dag}_{2,\uparrow}  \right)|0 \rangle_{e} \\
| \psi_{T3} \rangle & = & \hat{c}^{\dag}_{1, \downarrow} \hat{c}^{\dag}_{2,\downarrow} |0 \rangle_{e}.
\end{eqnarray}
The triplet states are not coupled
with the $b$-oscillator, and can be thus disregarded. The three other eigenstates of $\hat{H}_e$, denoted by $|-\rangle$, $|s+\rangle$ and $|s-\rangle$, are constructed from the singlet states and read explicitly
\begin{eqnarray}
| - \rangle & = & \frac{1}{\sqrt{2}} \left(  \hat{c}^{\dag}_{1, \uparrow} \hat{c}^{\dag}_{1,\downarrow}-\hat{c}^{\dag}_{2, \uparrow} \hat{c}^{\dag}_{2,\downarrow}  \right)|0 \rangle_{e} \\
|s+\rangle & = & \sin \theta |s \rangle-\cos \theta |+ \rangle \\
|s-\rangle & = & \cos \theta |s \rangle+\sin \theta |+ \rangle \\
\end{eqnarray}
where we have set
\begin{eqnarray}
| + \rangle & = & \frac{1}{\sqrt{2}} \left(  \hat{c}^{\dag}_{1, \uparrow} \hat{c}^{\dag}_{1,\downarrow}+\hat{c}^{\dag}_{2, \uparrow} \hat{c}^{\dag}_{2,\downarrow}  \right)|0 \rangle_{e} \\
| s \rangle & = & \frac{1}{\sqrt{2}} \left(  \hat{c}^{\dag}_{1, \uparrow} \hat{c}^{\dag}_{2,\downarrow}-\hat{c}^{\dag}_{1, \downarrow} \hat{c}^{\dag}_{2,\uparrow}  \right)|0 \rangle_{e}
\end{eqnarray}
and
\begin{equation}
\cos \theta= \sqrt{\frac{1}{2} \left( 1+\frac{U}{\sqrt{U^2+16 t^2}} \right)}.
\end{equation}
For these states one has $\hat{H}_e |-\rangle=U |-\rangle$ and $\hat{H}_e |s \pm \rangle=E_{\pm} |s \pm \rangle$, where
\begin{equation}
E_{\pm}=\frac{U \pm \sqrt{U^2+16t^2}}{2}.
\end{equation}
The {\it e-ph} interaction term $\hat{V}_{e-ph}$ couples the electronic eigenstates $|-\rangle$ and $| s \pm \rangle$ of $\hat{H}_e$ with the
phonon states $|n\rangle_{ph}=(1/ \sqrt{n!}) \hat{b}^{\dag n} |0 \rangle_{ph}$ of $\hat{H}_{ph}$.  \par

\subsection{Fock-space representation}
 In order to realize a photonic analog simulator of the HH Hamiltonian, it is worth
expanding the state vector $| \psi(t) \rangle$ of the electron-phonon system in Fock space as follows
\begin{equation}
| \psi_{t} \rangle = \sum_{n=0}^{\infty} \left[  f_n(t) |-\rangle +g_n |s + \rangle  -h_n(t) |s- \rangle  \right] \bigotimes |n \rangle_{ph}.
\end{equation}
The evolution equations for the amplitude probabilities $f_n$, $g_n$ and $h_n$ in Fock space, as obtained from the
Schr\"odinger equation $i \partial_t | \psi(t) \rangle= \hat{H} | \psi(t) \rangle$, read explicitly
\begin{eqnarray}
i \frac{df_n}{dt} & = & (U+\Omega n)f_n-\kappa_{n} g_{n-1}- \kappa_{n+1} g_{n+1}-\rho_n h_{n-1}+ \nonumber \\
& - & \rho_{n+1}h_{n+1} \\
i \frac{dg_n}{dt} & = & (E_++n \Omega)g_n-\kappa_n  f_{n-1}- \kappa_{n+1} f_{n+1} \\
i \frac{dh_n}{dt} & = & (E_-+n \Omega)h_n-\rho_n  f_{n-1}- \rho_{n+1} f_{n+1} \
\end{eqnarray}
($n=0,1,2,3,...$), where we have set
\begin{equation}
\kappa_n = \sqrt {2n} g \cos \theta   \; , \; \rho_n=\sqrt {2n} g \sin \theta
\end{equation}
and where we have assumed $f_{-1}=g_{-1}=h_{-1}=0$ for definiteness.

\section{Photonic analog simulator}

\subsection{Waveguide array design}

Equations (17-19) govern the temporal dynamics in Fock space of the
HH Hamiltonian, and their optical implementation can be achieved
using a simple {\it linear} optical network because the coupled
equations of the Fock-state amplitudes are {\it linear} equations.
Let us first notice that Eqs.(17-19) decouple into two sets of
equations involving  the former the amplitudes $\{
f_{2n}(t),g_{2n+1}(t),h_{2n+1}(t) \}$, the latter the amplitudes
$\{f_{2n+1}(t),g_{2n}(t),h_{2n}(t) \}$ ($n=0,1,2,3,...)$. A
schematic of the optical networks that provide analog simulators of
the two sets of equations are depicted in Fig.1. Their physical
implementation can be obtained by using  either arrays of coupled
optical waveguides with engineered coupling rates and propagation
constants, where the temporal evolution of the Fock-state amplitudes
is mapped into spatial propagation of the light fields along the
axis of the various waveguides, or arrays of coupled optical
cavities with engineered coupling rates and resonance frequencies,
in which the temporal evolution of the Fock-state amplitudes is
mimicked by the temporal evolution of the light field stored in the
various cavities. Owing to the current technological advances in
manufacturing engineered arrayed waveguides and  related imaging
techniques based on  femtosecond laser writing (see, for instance,
\cite{Szameit10} and references therein), we will consider here
specifically an optical analog simulator of the HH model based on
propagation of monochromatic light waves in coupled optical
waveguides. Moreover, since the dynamics of the two networks of Fig.1 is
decoupled, without loss of generality we can limit to consider
initial conditions that excite only one of them, for example the one
corresponding to Fig.1(a).
\par
Light transport of a {\it  classical} monochromatic field at wavelength $\lambda$ in a weakly-guiding arrays of coupled optical waveguides with a straight optical axis $z$, in the geometrical setting schematically depicted in Fig.1(a),
is described  by the optical Sch\"{o}dinger equation for the electric field envelope $\phi(x,y,z)$ \cite{LonghiLPR}
\begin{equation}
i \lambdabar \frac{\partial \phi}{\partial z} = -\frac{\lambdabar^2}{2n_s} \left( \frac{\partial^2 \phi}{\partial x^2} +  \frac{\partial^2 \phi}{\partial y^2} \right)+V(x,y) \phi,
\end{equation}
 where $\lambdabar= \lambda/(2 \pi)$ is the reduced wavelength of photons, $n_s$ is the refractive index of the substrate at wavelength $ \lambda$, $V(x,y) \simeq n_s-n(x,y)$ is
 the so-called optical potential, and $n(x,y)$ is the refractive index profile of the optical structure defining the various waveguide channels. Let  $h(x,y)$ be the normalized profile of the waveguide core, which is typically assumed to be  circularly-shaped with a super-Gaussian profile of radius $a$, $(x_i,y_i)$ the position of the $i$-th waveguide of the array, and $\Delta n_i$ its index change from the cladding (substrate) region. One can thus write:
 \begin{equation}
 V(x,y) \simeq -\sum_{i} \Delta n_i h(x-x_i,y-y_i).
 \end{equation}
 In the tight-binding
approximation, the electric field envelope can be expanded as a superposition of the fundamental guided modes of each waveguide with
 amplitudes that vary slowly along $z$ due to evanescent mode-coupling and propagation constant mismatch \cite{Yariv}. In the nearest-neighboring approximation, the resulting coupled-mode equations, obtained by using e.g. a variational procedure  \cite{LonghiPLA}, are precisely of the form (17-19), provided that cross-couplings among waveguides are negligible and the temporal variable $t$ in the quantum HH model is replaced by the spatial propagation distance $z$. If needed, to avoid cross-couplings suitable high-index waveguides (or off-resonance resonators) could be inserted into the structure, as shown  in Fig.1(a).
 The coupling constants $\kappa_n$, $\rho_n$ entering in the equations (17-19) are basically determined by the waveguide separation via an overlapping integral of the evanescent fields \cite{LonghiPLA}, whereas the propagation constant shifts $U+n \Omega$ and $E_{\pm}+n \Omega$ can be controlled by the core index changes $\Delta n_n$ of the various waveguides.  Let us consider, as an example, propagation of light waves at wavelength $\lambda=633$ nm (red) in waveguides written by femtosecond laser pulses in fused silica \cite{Szameit10}, for which $n_s \simeq 1.45$. The normalized core profile of the waveguides is assumed to be super-Gaussian-shaped with a core radius $a=2.5 \; \mu$m, namely $h(x,y)=\exp[-(x^2+y^2)^2/a^4]$; the index change of the various waveguides is taken to be $\Delta n_i=\Delta n+\delta n_i$, where $\Delta n=2 \times 10^{-3}$ is a reference index change and $\delta n_i$ a small perturbation, as compared to $\Delta n$, needed to realize the propagation constant shifts. For such waveguides, the coupling constant $\kappa(d)$ (in units of ${\rm mm}^{-1}$) between two waveguides placed at a distance $d$ (in units of $\mu$m) turns out to be very well approximated, for distances $d > \sim 6 \mu$m, by the exponential relation \cite{LonghiPRB2009} $\kappa=A \exp(-\gamma d)$, with $A \simeq 24.6 \; {\rm mm}^{-1}$ and
 $\gamma \simeq 0.466 \; \mu {\rm m}^{-1}$. Similarly, the propagation constant shift $\sigma(\delta n)$ of the fundamental waveguide mode induced by an index change $\delta n$, added to $\Delta n$, is very well approximated (for $|\delta n| \ll \Delta n$) by the relation $\sigma(\delta n) \simeq -1.7 \delta n / \lambdabar$.  Using such relations, one can thus design a waveguide array, with controlled waveguide separation and index changes, that realizes  the desired coupling rates $\kappa_n$, $\rho_n$ [given by Eq.(20)] and propagation constant shifts $U+n \Omega$ and $E_{\pm}+n \Omega$. The total number of waveguides needed to simulate the HH model is basically determined by the largest phonon number $n$ that will be excited in the dynamics.
\begin{figure}
\includegraphics[scale=0.7]{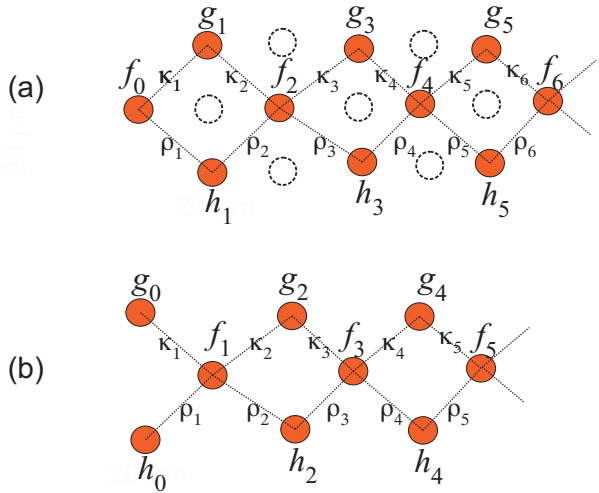}
\caption{Schematic of the optical networks that realize analog
simulators of the two sets of equations (17-19) for $n$ even
[Fig.1(a)] and $n$ odd [Fig.1(b)]. The filled circles indicate either optical waveguides or optical cavities (resonators) that are coupled by
weak evanescent field. The optical elements depicted in Fig.1(a) by  dotted circles 
indicate  high-index core waveguides (or off-resonance cavities) that might be inserted, if needed, to avoid cross coupling 
among the optical elements in the network.}
\end{figure}

\begin{figure}
\includegraphics[scale=0.42]{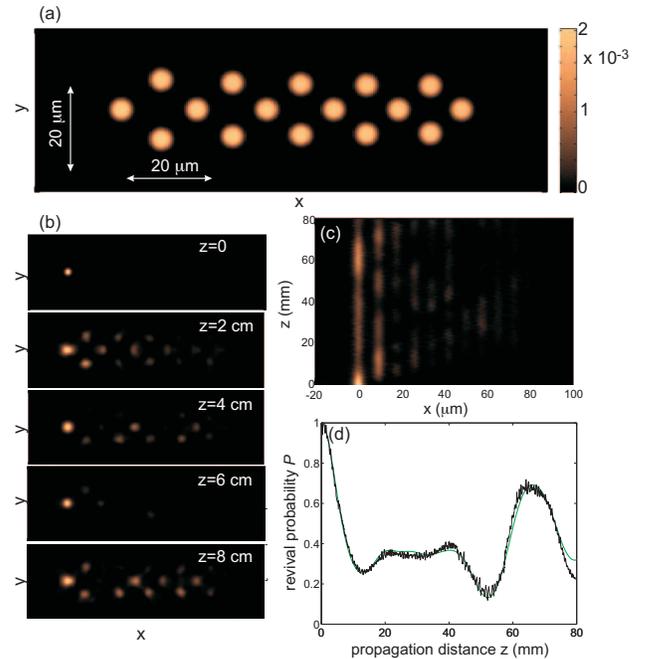}
\caption{Optical simulations of the HH model in Fock space for
parameter values $g=0.1 \; \rm{mm}^{-1}$, $t=0.1\; \rm{mm}^{-1}$,
$\Omega=0.1 \; \rm{mm}^{-1}$, and $U=0$ (uncorrelated electrons). (a)
Refractive index profile $n(x,y)-n_s$ of the double-chain zig-zag waveguide array. (b-d)
Dynamical evolution of the system corresponding to the initial state
$|\psi(0) \rangle=|- \rangle \bigotimes |0 \rangle_{ph}$. In (b) a
snapshot of the intensity light pattern $|\phi(x,y,z)|^2$ is shown
for a few propagation distances $z$, whereas in (c) a snapshot of
$\int dy |\phi(x,y,z)|^2$ in the $(x,z)$ plane is depicted. Figure
(d) shows the behavior of the return probability $P(z)=|\langle
\psi(0)|\psi(z) \rangle|^2=|f_0(z)|^2$ numerically obtained from the
solution of the optical Schr\"{o}dinger equation (solid curve) and
from the HH model (thin solid curve). }
\end{figure}

 \subsection{Numerical results}
 In this section we provide a few examples of photonic simulations of the temporal dynamics of the HH Hamiltonian in Fock space, based on light transport
  in waveguide arrays engineered following the procedure outlined in the previous subsection, and compare the numerical results obtained by the
  full numerical analysis the optical Schr\"{o}dinger equation (21) with those predicted by the coupled-mode equations (17-19).
  As an initial condition, we will assume the system to be in the state $| \psi(0) \rangle =|- \rangle  \bigotimes |0\rangle_{ph} $, corresponding to
   excitation of zero phonon modes and to the two electrons being in the entangled state
   $|-\rangle=(1/ \sqrt{2}) ( \hat{c}^{\dag}_{1, \uparrow} \hat{c}^{\dag}_{1,\downarrow}-\hat{c}^{\dag}_{2, \uparrow} \hat{c}^{\dag}_{2,\downarrow} ) |0 \rangle$,
   i.e. the two electrons occupying  the same (either left or right) potential well with opposite spins.
 For such an initial state, according to Eq.(16) the initial conditions for the Fock-state amplitudes are given by
 $f_n(0)=\delta_{n,0}$ and $g_n(0)=h_n(0)=0$. Such an initial condition simply corresponds to the excitation of the
   left boundary waveguide in the semi-array of Fig.1(a). The fractional light powers trapped in the various waveguides of the array
at different propagation distances $z$ provide a direct mapping of
the temporal evolution of the occupation probabilities $|f_n|^2$, $|g_n|^2$ and
$|h_n|^2$ of the quantum HH Hamiltonian in Fock space.
   To monitor the temporal dynamics of the HH model, we mainly use as a dynamical variable
    the return probability $P(t)=| \langle \psi(0)|\psi(t) \rangle |^2 =|f_0(t)|^2$. In our optical simulator, such a variable
    simply corresponds to the fractional light power that remains in the
initially excited boundary waveguide. In the simulations, we assumed
constant values of ${\it e-ph}$ coupling $g=0.1 \; \rm{mm}^{-1}$
 and hopping rate $t=0.1\;
\rm{mm}^{-1}$, whereas a few values of the on-site ${\it e-e}$
Coulomb energy $U$ are considered, corresponding to three
different correlation regimes: absence of correlation ($U=0$),
moderate correlation regime ($U$ comparable with the hopping rate
$t$); and strong-correlation regime ($U$ much larger than $t$)
\cite{note}. \par
\begin{figure}
\includegraphics[scale=0.42]{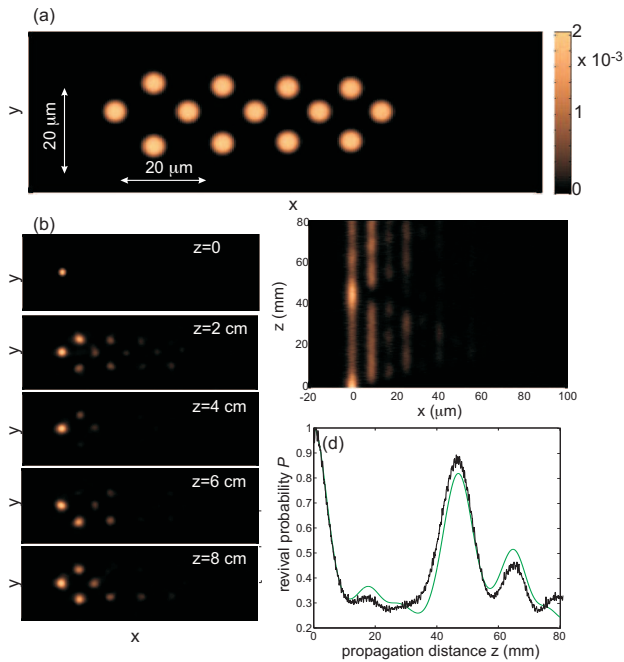}
\caption{Same as Fig.2, but in the intermediate correlation regime
($U=0.2 \; \rm{mm}^{-1}$).}
\end{figure}
\begin{figure}
\includegraphics[scale=0.42]{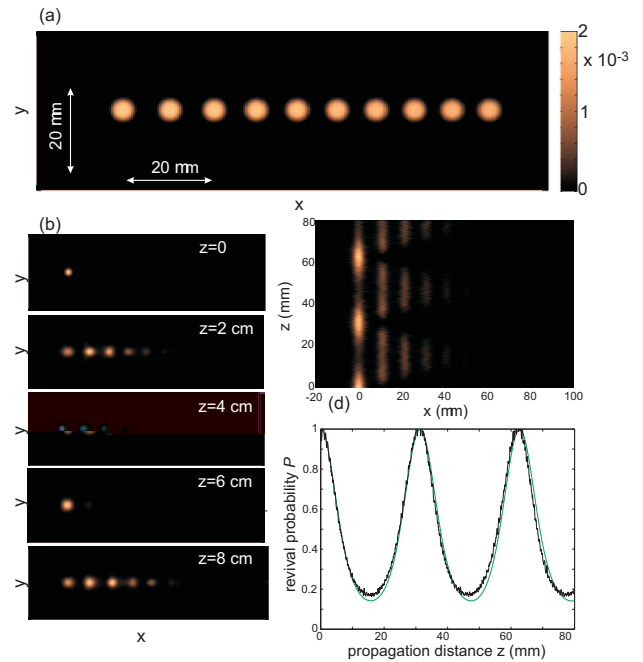}
\caption{Same as Fig.2, but in the strong correlation regime ($U=10
\; \rm{mm}^{-1}$). The oscillation frequency is now $\Omega=0.2 \;
\rm{mm}^{-1}$.}
\end{figure}

 Figure 2 shows the results obtained  for uncorrelated electrons ($U=0$) and for an oscillator frequency
$\Omega=0.1 \; \rm{mm}^{-1}$. In this case, one has $\cos
\theta=\sin \theta=1/\sqrt 2$, and thus $\kappa_n=\rho_n$. The refractive index profile
$[n(x,y)-n_s]$ of the waveguide array is depicted in Fig.2(a). The array comprises
$16$ waveguides arranged in two zig-zag chains with non-uniform spacings. The evolution of the light intensity along the
array is shown in Figs.2(b) and (c). In particular, in Fig.2(b) a
snapshot of the intensity light distribution $|\phi(x,y,z)|^2$ is
depicted for a few propagation distances (up to $z=8$ cm) as
obtained by numerical simulations of the optical Schr\"{o}dinger
equation (21), whereas in Fig.2(c) the snapshot of $\int  dy
|\phi(x,y,z)|^2$ in the $(x,z)$ plane is shown. Note that a
relatively small number of waveguides ($\sim 13$) are excited in the
dynamics. The corresponding behavior of the return probability
$P(z)$ is depicted in Fig.2(d) (solid curve) and compared with the
behavior obtained by direct numerical analysis of the HH equations
(17-19) (thin solid curve). Note that, even though in our optical network high-index waveguides to avoid weak 
cross couplings have not been inserted, the agreement with the HH model [Eqs.(17-19)] is rather satisfactory.  
 Figure 3 shows similar results obtained in the moderate correlated regime ($U=0.2 \; {\rm
 mm}^{-1}$), corresponding to $\cos \theta \simeq 0.851$, $\sin \theta
 \simeq 0.526$. Note that in this case $\kappa_n \neq \rho_n$, and thus the two zig-zag chains forming the array are
 noticeably asymmetric.\\ 
 The strong correlation
 regime is attained when the on-site interaction strength $U$ is
 much larger than the hopping amplitude $t$, corresponding to $\cos \theta
 \simeq 1$ and $\sin \theta \simeq 0$. In this case $\rho_n \simeq
 0$ and thus the phonon modes are coupled to the electronic degree
 of freedoms solely via the two dressed states $|- \rangle$ and
 $|s+\rangle \simeq -|+\rangle$, whose energy levels are nearly degenerate and equal to
 $U$ and $E_{+} \simeq U(1+4t^2/U^2)$, respectively. In Fock space, the ${\it e-ph}$ coupling dynamics is thus reduced to the coupled equations for
 the occupation amplitudes $f_n$ and $g_n$,
 namely [see Eqs.(17) and (18)]
\begin{eqnarray}
i \frac{df_n}{dt} & = & (U+\Omega n)f_n-\kappa_{n} g_{n-1}-
\kappa_{n+1} g_{n+1} \\
 i \frac{dg_n}{dt} & = & (E_++n
\Omega)g_n-\kappa_n  f_{n-1}- \kappa_{n+1} f_{n+1}
\end{eqnarray}
where $k_n=\sqrt{2n} g$.
 Correspondingly, the optical analog simulator reduces to a simple linear chain of waveguides with non-uniform spacing, as shown in Fig.4(a).
  An example of optical simulation of the HH model
 in the strong correlation regime is shown in Fig.4 for parameter values
 $g=0.1 \; \rm{mm}^{-1}$, $t=0.1\; \rm{mm}^{-1}$, $\Omega=0.2 \; \rm{mm}^{-1}$
and $U=10 \; \rm{mm}^{-1}$. Note that the dynamics of occupation
amplitudes shows in this case a typical periodic behavior with a
period given by $2 \pi / \Omega$. In our optical simulator, the
periodic revival is clearly visualized by the self-imaging property
of the waveguide array, in which the initial light periodically
returns into the initially-excited waveguide [see Fig.4(c)]. Such a
periodic behavior can be explained after observing that, in the
strong correlation  regime $U \rightarrow \infty$, the phonon modes
$|n \rangle_{ph}$ couple solely to {\it two} dressed electronic
states $|- \rangle$ and $|+ \rangle$. Neglecting the small energy
shift $(E_+ -U)$ of dressed levels $|- \rangle$ and $|+ \rangle$, the
energy levels of the ${\it e-ph}$ coupled system [the eigenvalues of Eqs.(23) and (24)
with $E_+=U$]  is analogous to a Wannier-Stark ladder  with energy spacing $\Omega$, namely one has $E_l=U+l \Omega-2g^2/\Omega$ ($l=0,1,2,3,...$). In fact, in this
limit the HH model is equivalent to a two-site tunneling system
strongly coupled to a quantum harmonic oscillator, such as the
two-site Holstein model in atomic (strong ${\it e-ph}$ coupling)
limit \cite{H}, describing the hopping dynamics of {\it one}
electron in a two-site potential interacting with local phonon
modes, or the Jaynes-Cummings model of quantum optics, describing
the interaction of a two-level atom with a quantized oscillator in the
so-called deep strong coupling regime \cite{JC}. In the atomic
(strong-coupling) limit, i.e. for a vanishing hopping rate, the
Holstein model is exactly integrable using a Lang-Firsov
transformation \cite{HH0,Fir}, yielding an equally-spaced discrete energy spectrum
corresponding to small Holsten polaron modes .
In our optical realization, the equally-spaced energy spectrum is
responsible for the periodic self-imaging of the optical beam as it
propagates along the array, with a spatial period $\ 2 \pi/ \Omega$.
In our realization of the HH model in Fock space, the periodic self-imaging effect 
can be explained in terms of generalized 
{\it Bloch oscillations} of a {\it single} particle  in a {\it non-homogeneous}
tight-binding lattice with a dc bias. In fact, the tight-binding lattice
realization of the HH model in the strong correlation regime, as
described by Eqs.(23) and (24) with $E_+ \simeq U$, belongs to a
class of exactly-solvable lattice models which shows an
equally-spaced Wannier-Stark ladder spectrum, as shown in
Ref.\cite{LonghiSELF}. In this lattice model, the dc biased in represented by the linear gradient terms $n \Omega$ in Eqs.(23) and (24), and 
Bloch oscillations are not washed out in spite of lattice truncation and non-homogeneity of hopping rates Ref.\cite{LonghiSELF}..
If the small energy shift between the dressed
energy states $|- \rangle$ and $|+ \rangle$ is taken into account,
the revival (self-imaging) dynamics would be only approximate
because the energy levels of $\hat{H}$ are no more exactly equally
spaced (see, for instance, \cite{JC}). However, for the parameter
values used in the simulation of Fig.4, washing out of the self-imaging would be
visible after several self-imaging periods, and it is thus not
observed in the figure.

\section{Conclusions}
In this work, a classical analog simulator of the Hubbard-Holstein model, describing electron-phonon coupling in presence of electron correlations, has been theoretically proposed based on light transport in engineered optical waveguide lattices. Here a toy model has been presented by considering a diatomic molecule, i.e. the two-site two-electron HH model, which describes two correlated electrons hopping between two adjacent sites. Like quantum simulators based on single photons in linear optical networks \cite{ppp2}, the kind of classical simulators proposed here is suited to mimic few interacting particles solely, rather than many-particles like quantum simulators based on trapped atoms or ions. Nevertheless, our classical simulator offers the rather unique possibility to  visualize the temporal dynamics of the quantum system in Fock space (rather than in real space) and to mimic dynamical evolutions of quite arbitrary and controllable initial states, such as entangled states, which could be not accessible in condensed-matter systems. Simulation of  the HH model in Fock space also helps to get new physical insights into certain dynamical aspects of the HH model.
For instance, in the strong correlation regime the periodic temporal dynamics exhibited by the HH Hamiltonian, related to the excitation of Holstein polarons, can be explained in terms of generalized Bloch oscillations of a single particle in a semi-infinite inhomogeneous tight-binding lattice, a result which was not noticed in previous studies of HH models. It is envisaged that the present work could stimulate the search for and the experimental demonstration of further {\it classical} simulators of many-body problems, involving  other types of interactions and even beyond the two-site toy model. For example, light transport in two-dimensional photonic lattices can simulate in Fock space the hopping dynamics of two correlated electrons on a one-dimensional lattice, and can be exploited to visualize such phenomena as correlated tunneling of bound electron-electron molecules and their coherent motion under the action of dc or ac fields, such as Bloch oscillations and dynamic localization of strongly correlated particles \cite{ex1}.   

 \acknowledgments
 The authors acknowledge financial support by the
Italian MIUR (Grant No. PRIN-2008-YCAAK project "Analogie
ottico-quantistiche in strutture fotoniche a guida d'onda").


\begin{thebibliography}{}
\bibitem{note}
Since in our optical analog simulator the temporal variable $t$ of
the quantum problem is mapped into the spatial propagation distance
$z$ along the waveguide array, the parameters $t$, $g$, $\Omega$ and
$U$ entering in the HH Hamiltonian are given in units of a spatial
frequency.
\end{thebibliography}
\end{document}